\documentclass[10pt,twocolumn,oneside,letterpaper,conference]{IEEEtran}
\IEEEoverridecommandlockouts
\usepackage{cite}
\usepackage{amsmath,amssymb,amsfonts}
\usepackage{graphicx}
\usepackage{textcomp}
\usepackage{xcolor}
\usepackage{xcolor}

\usepackage{cite}
\usepackage{graphicx}
\usepackage{epstopdf}

\usepackage{amsmath}
\usepackage{times}
\usepackage{latexsym}
\usepackage{graphicx}
\usepackage{bm}
\usepackage{amssymb}
\usepackage{caption}
\usepackage{stfloats}
\usepackage{cases}
\usepackage{array}
\usepackage{setspace}
\usepackage{fancyhdr}
\usepackage{fancyhdr}
\usepackage{algorithm}
\usepackage{algorithmicx}
\usepackage{algpseudocode}
\usepackage{textcomp}
\usepackage{wasysym}
\usepackage{balance} %balance package 1
\usepackage{flushend} %balance package 2
\usepackage{url}
\usepackage{xcolor}
\usepackage{subfigure}
\usepackage{stfloats}
\usepackage{enumerate}
\usepackage{array}
\usepackage{xcolor}
\usepackage{tikz}
\usepackage{booktabs}
\usepackage{colortbl}

\def\BibTeX{{\rm B\kern-.05em{\sc i\kern-.025em b}\kern-.08em
    T\kern-.1667em\lower.7ex\hbox{E}\kern-.125emX}}

\begin{document}

\title{Social-aware Cooperative Caching in Fog Radio Access Networks}

\author{
\IEEEauthorblockN{Baotian Fan$^{1}$,
Yanxiang Jiang$^{1,2,*}$, Fu-Chun Zheng$^{1,2}$, Mehdi Bennis$^3$, and~Xiaohu~You$^1$}
%and Xiaohu You$^1$}
\IEEEauthorblockA{$^1$National Mobile Communications Research Laboratory,
Southeast University, Nanjing 210096, China\\
$^2$School of Electronic and Information Engineering, Harbin Institute of Technology, Shenzhen 518055, China\\
$^3$Centre for Wireless Communications, University of Oulu, Oulu 90014, Finland\\
E-mail: $\{$220180743@seu.edu.cn, yxjiang@seu.edu.cn, fzheng@ieee.org, mehdi.bennis@oulu.fi, xhyu@seu.edu.cn$\}$}} %, xhyu@seu.edu.cn $\}$

\maketitle

\begin{abstract}
In this paper, the cooperative caching problem in fog radio access networks (F-RANs) is investigated to jointly optimize the transmission delay and energy consumption. Exploiting the potential social relationships among fog access points (F-APs), we firstly propose a clustering scheme based on hedonic coalition game (HCG) to improve the potential cooperation gain. Then, considering that the optimization problem is non-deterministic polynomial hard (NP-hard), we further propose an improved firefly algorithm (FA) based cooperative caching scheme, which utilizes a mutation strategy based on local content popularity to avoid  pre-mature convergence. Simulation results show that our proposed scheme can effectively reduce the content transmission delay and energy consumption in comparison with the baselines.
\end{abstract}


\section{Introduction}
With the rapid popularization of intelligent mobile terminals and the continuous development of application services, wireless networks are undergoing tremendous data traffic pressure. The unprecedented data traffics especially bring huge traffic congestion to the fronthaul links. Fog radio access network (F-RAN) is regarded as a promising network architecture to effectively alleviate traffic congestion and reduce content delivery delay by deploying storage and computing resources in fog access points (F-APs) at the network edge\cite{Peng,8632748,jiang2020analysis,peng2018performance, feng2019content, lu2019distributed, hu2018distributed}. To fully exploit the limited caching capacity, cooperative caching becomes a viable approach to decrease energy consumption and content delivery delay.

Recently, many studies have been carried out on cooperative caching.
In \cite{Xia}, a cooperative edge caching scheme was proposed  to minimize the average download delay by exploiting an improved pigeon inspired optimization (PIO) approach.
In \cite{Fang}, a cooperative caching algorithm was proposed for clustered vehicular content networks, which was devised based on global caching status.
In \cite{cuix}, a cooperative content placement problem was investigated to maximize the increasing offloaded traffic by utilizing a graph-based approach.
In \cite{ZhangShan}, by considering the stochastic information of network topology, a user-centric cooperative caching scheme was proposed to minimize the average content transmission delay.
In \cite{ZhongC}, based on multi-agent reinforcement learning, a cooperative content strategy was proposed to maximize the average transmission delay reduction without the knowledge of user preference in advance.
However, most of the above mentioned research works give several fixed clusters in advance, and do not form clusters through potential relationships among access points.
Besides, the content diversity in access points will be further decreased without consideration of cooperation among multiple clusters.
%\textcolor{blue}{Moreover, in \cite{Xia,Fang,Cui}, the basic parameters in the models are too small to reveal the cache performance of the schemes.}
%non-cooperation among multiple clusters further reduced the diversity of content in F-APs.

Motivated by the aforementioned discussions,
we propose a cooperative caching scheme to jointly optimize the transmission delay and energy consumption.
We firstly utilize local content popularity, distance and transmission power of F-APs to measure their social relationships. Then, we propose a social-aware clustering scheme by exploiting hedonic coalition game (HCG) to improve the cooperation gain among F-APs.
Due to the non-deterministic polynomial hard (NP-hard) property of the optimization problem, we further propose an improved firefly algorithm (FA) based cooperative caching scheme, which can obtain a better feasible solution with low computational complexity.

%Motivated by the aforementioned discussions, we utilize local content popularity, distances and transmission power of F-APs to measure the hidden social relationships among F-APs. Meanwhile, we propose a social-aware-based clustering approach by exploiting hedonic coalition game (HCG) to obtain more cooperation gains and improve the diversity of contents obtained from F-APs. \textcolor{blue}{The cooperative caching optimization problem is formulated to jointly optimize the transmission delay reduction and energy consumption.}Due to the non-deterministic polynomial hard property (NP-hard) of the optimization problem, we propose an improved firefly algorithm (FA) \textcolor{blue}{which can} decrease the computational complexity of finding the global optimum.

The rest of this paper is organized as follows.
In Section II, the system model and problem formulation are briefly presented.
In Section III, the proposed HCG-based clustering scheme is introduced.
The improved FA-based cooperative caching scheme is presented in Section IV.
Simulation results are shown in Section V.
Conclusions are drawn in Section VI.

%\newpage
\section{System Model and Problem Formulation}
\subsection{Network Model}

\begin{figure}[!t]
	\centering %\vspace*{135pt}
	\includegraphics[width=0.30\textwidth]{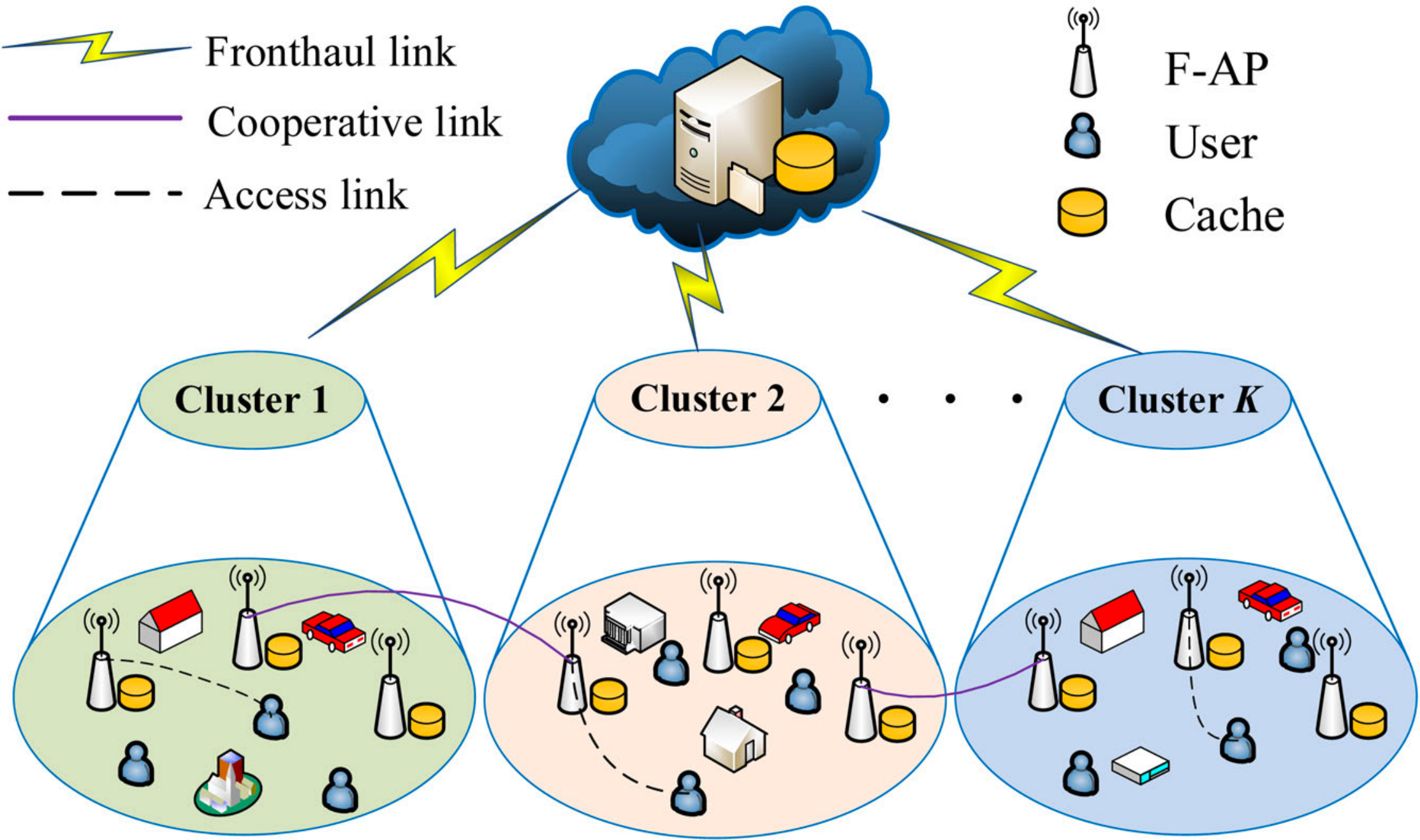}
	%\captionstyle{mystyle3}
	\caption{Illustration of the cooperative caching scenario in F-RANs.}
	\label{model}
\end{figure}

As shown in Fig. \ref{model}, we consider a specific cooperative caching  region in F-RANs, which consists of a cloud server, $M$ F-APs and $U$ users.
The cloud server can collect relevant caching status information of the F-RANs, and broadcast the information to all F-APs in the considered region \cite{LiZ}.
Let $\mathcal{M}=\{1, 2,\ldots, m, \ldots, M\}$ and $\mathcal{U}=\{1, 2,\ldots, u, \ldots U\}$ denote the F-AP set and the user set, respectively.
%For the sake of simplicity, i
If user $u$ sends a content request to the nearest F-AP $m$, then F-AP $m$ is referred to as the local F-AP of user $u$ and user $u$ is referred to as the local user of F-AP $m$. Let $\mathcal{U}_{m}$ denote the set of local users for F-AP $m$.
By considering the intensive deployment of F-APs, they can cooperate with each other and form several disjoint clusters.
%to reduce content delivery delay and energy consumption
%by exploiting the potential social relationships among F-APs.
Suppose that the considered F-APs can be divided into $K$ disjoint clustered sets denoted as $\mathcal{S}_{k}$ for $k \in \{1, 2, \ldots,k, \ldots, K\}$.
Then, we have: $\mathcal{M}=\bigcup_{k=1}^{K} \mathcal{S}_{k}$, $\mathcal{S}_{k} \bigcap \mathcal{S}_{k^{\prime}}=\varnothing$, $\forall \mathcal{S}_{k}, \mathcal{S}_{k^{\prime}} \subset \mathcal{M},{k} \ne {k'}$.
%let $\mathcal{S}_{k}$ denote a cluster formed by several F-APs with social relationships. Let $\{\mathcal{S}_{1}, \mathcal{S}_{2}, \ldots, \mathcal{S}_{k}, \ldots \mathcal{S}_{K}\}$ denote the set of $K$ disjoint clusters.

Without loss of generality, we assume that all the contents, which come from the content library $\mathcal{F}=\{1, 2,\ldots, f, \ldots F\}$ located in the cloud server, have the same size $L$. Let $C$ denote the storage capacity of each F-AP,
$p_{u, f}$ the probability of user $u$ requesting for content $f$, and $p_{m,f}$ the probability that all local users of F-AP $m$ send requests for content $f$ (referred to as the normalized local content popularity of content $f$ at F-AP $m$). Then, we have:
\begin{equation}
p_{m, f}=\frac{\sum_{u \in \mathcal{U}_{m}} p_{u, f}}{\sum_{f \in \mathcal{F}} \sum_{u \in \mathcal{U}_{m}} p_{u, f}}.
\end{equation}

If the requested content has been stored in local F-AP of the same cluster, it will be directly transmitted to the requesting user via access link.
If the requested content has been stored in other clusters, it will be delivered to local F-AP via cooperative link and forwarded to the requesting user by local F-AP.
If the requested content has only been cached in the cloud server, local F-AP will obtain the content from the cloud server via fronthaul link and forward to the requesting user.
%they will be delivered to local F-APs from the cloud server via fronthaul links and then forwarded to the requesting users.
%the users will obtain the contents from the cloud server via fronthaul links.

In addition, in our network model, the content caching process is composed of two phases: content placement and content delivery. Since content placement phase usually occurs at off-peak time, it does not occupy content delivery time.
% \cite{Huang}
\subsection{Content Delivery}
Let $R$ denote the average transmission rate of contents fetched from the cloud server.
Let $r_{m, u}$ denote the distance between F-AP $m$ and user $u$, and $r_{m, n}$ the distance between F-AP $m$ and F-AP $n$.
Let $B_{1}$ and $B_{2}$ denote the transmission bandwidths of access link and cooperative link, respectively.
Let $P_{m}$ denote the transmission power of F-AP $m$, $\sigma^{2}$ the noise power, $I$ the sum of interfering signal power from neighboring F-APs, and ${\alpha}$ the pathloss exponent.
Let $R_{m, u}$ denote the achievable content transmission rate of the access link between F-AP $m$ and user $u$. Then, it can be expressed as
$
R_{m, u}=B_{1} \log _{2}\left(1+{P_{m} r_{m, u}^{-\alpha}}/\left({\sigma^{2}}+{I}\right)\right)
$.
Let $R_{m, n}$ denote the achievable content transmission rate of the cooperative link between F-AP $m$ and F-AP $n$. Then, it can be expressed as
$
R_{m, n}=B_{2} \log _{2}\left(1+{P_{m} r_{m, n}^{-\alpha}}/\left({\sigma^{2}}+{I}\right)\right)
$.

Let $x_{m, f}$ denote the binary cache decision variable with $x_{m, f}=1$ if F-AP $m$ has cached content $f$ and  $x_{m, f}=0$ otherwise.
For the sake of simplicity, we define $x_{k, f}=1-\prod_{m \in \mathcal{S}_{k}}\left(1-x_{m, f}\right)$ \cite{cuix}. Specifically, if cluster $S_{k}$ has cached content $f$, we have $x_{k, f}=1$. Otherwise, $x_{k, f}=0$.

If user $u$ requests content $f$ which has been stored in local cluster $\mathcal{S}_{k}$, the transmission delay %\cite{ZhongC}
 can be expressed as follows:
\begin{equation}
T_{1}= x_{k, f}{L}/{R_{m, u}}.
\end{equation}
If the requested content has not been cached in local cluster $\mathcal{S}_{k}$ and has been stored in another cluster $\mathcal{S}_{k^{\prime}}$, the transmission delay can be expressed as follows:
\begin{multline}
	T_{2}=\left(1-x_{k, f}\right)\left[1-\prod\nolimits_{\mathcal{S}_{k^{\prime}} \subset \mathcal{M}}\left(1-x_{k^{\prime},f} \right)\right]\\
	\times \left(1/{R_{m, n}}+1/{R_{m, u}}\right){L}.
\end{multline}
If the requested content has only been cached in the cloud server, the transmission delay can be expressed as follows:
\begin{multline}
	T_{3}=\left(1-x_{k, f}\right)\prod\nolimits_{\mathcal{S}_{k^{\prime}} \subset \mathcal{M}}\left(1-x_{k^{\prime}, f}\right) \\
	\times \left(1/{R}+1/{R_{m, u}}\right){L}.
\end{multline}
Then, according to \cite{Xia}, the average transmission delay for obtaining the requested contents can be expressed as follows:
\begin{equation}
T=\sum_{u\in \mathcal{U}} \sum_{f\in \mathcal{F}} p_{u, f}\left(T_{1}+T_{2}+T_{3}\right).
\end{equation}

\subsection{Energy Consumption}
We adopt the energy consumption model in \cite{LiuD}. %with consideration of caching energy consumption and transmission energy consumption.
Let $E_{\mathrm{c}}$ denote the energy consumption for content caching. Then, it can be expressed as follows:
\begin{equation}
E_{\mathrm{c}}=\sum_{m=1}^{M} \sum_{f=1}^{F} x_{m, f} J_{\mathrm{c}} {L},
\end{equation}
where $J_{\mathrm{c}}$ is the power coefficient of cache hardware.

When content $f$  is transmitted to the requesting user $u$, the energy consumption is divided into three parts. If the requested content has been stored in local cluster $\mathcal{S}_{k}$, the energy consumption for content delivery can be expressed as follows:
\begin{equation}
E_{1}=x_{k, f}{L} P_{m}/{R_{m, u}}.
\end{equation}
If the requested content has been stored in another cluster $\mathcal{S}_{k^{\prime}}$, the energy consumption for content delivery can be expressed as follows:
\begin{multline}
	E_{2}=\left(1-x_{k, f}\right)\left[1-\prod\nolimits_{\mathcal{S}_{k^{\prime}} \subset \mathcal{M}}\left(1-x_{k^{\prime},f} \right)\right]\\
	\times \left({1}/{R_{m, n}}+{1}/{R_{m, u}}\right) P_{m}{L}.
\end{multline}
If the requested content has only been cached in the cloud server, the energy consumption for content delivery can be expressed as follows:
\begin{multline}
	E_{3}=\left(1-x_{k, f}\right)\prod\nolimits_{\mathcal{S}_{k^{\prime}} \subset \mathcal{M}}\left(1-x_{k^{\prime}, f}\right)\\
	\times \left(P_{\mathrm{s}}/{R}+P_{m}/{R_{m, u}}\right){L},
\end{multline}
where $P_{\mathrm{s}}$ denotes the average transmission power of contents fetched from the cloud server.

Correspondingly, the system energy consumption for obtaining the requested contents can be expressed as follows:
\begin{equation}
E=E_{\mathrm{c}}+\sum_{u\in \mathcal{U}} \sum_{f\in \mathcal{F}} p_{u, f}\left(E_{1}+E_{2}+E_{3}\right).
\end{equation}

\subsection{Problem Formulation}
To minimize the transmission delay and energy consumption, the corresponding cooperative caching optimization problem can be formulated as follows:
\begin{align}\label{P-1}
	&{\min_{x_{m, f}}    \mu T+(1-\mu) E}, \\
	{\text{s.t.}}\
	&{\sum_{f=1}^{F} x_{m, f} {L} \leq C,\forall m \in \mathcal{M}},\tag{\ref{P-1}a} \nonumber\\
	&{x_{m, f} \in\{0,1\}, \forall m \in \mathcal{M},\forall f \in \mathcal{F}}, \tag{\ref{P-1}b}\nonumber
\end{align}
where $\mu$ is the scalar weight to ensure the same range for the transmission delay and energy consumption.
%Besides, (\ref{P-1}a) and (\ref{P-1}b) guarantee the constraints of storage space and integer, respectively.

\section{Proposed Social-aware Cooperative Caching Scheme}
The optimization problem in (\ref{P-1}) has two-fold challenges.
First, the formation of clusters is unknown and giving fixed clusters in advance does not improve the caching performance. %Hence, in order to improve the caching performance and accelerate the problem solving, it is necessary to optimize the formation of clusters.
Second, the optimization problem
%is a mixed integer programing problem and
%is NP-hard. Hence, it is necessary to reduce the complexity to efficiently solve the problem.
has the NP-hard property and exponential computational complexity is required for finding the optimal solution.
%Hence, it is necessary to solve the optimization problem in polynomial time.
To overcome the two-fold challenges, in the following subsections, we solve the optimization problem by optimizing the formation of clusters and content placement separately.

\subsection{Proposed HCG-based Clustering}
In order to optimize the formation of clusters, we firstly measure the social relationships among F-APs and define the social preferences of F-APs for clusters. Then, we model the formation of clusters as a hedonic coalition game and propose a HCG-based clustering scheme.

\subsubsection{Social Preference}
%\subsubsection{Social Preference}
Social networks are theoretical constructions composed of units and ties, which have been widely applied to study the relationships among groups or communities. F-APs can be considered as social units and the relationships among F-APs can be considered as ties \cite{ZHANGX}.
% An initial question is how to  measure the social relationships among the F-APs.
Inspired by \cite{LIY}, we extend the social relationships among users to F-APs.

%\begin{comment}
%Without loss of generality, we assume that the corresponding contributions could be made by other F-APs, when other F-APs assist F-AP $m$ in transferring content $f$.
%If the content $f$ has a higher popularity in F-AP $m$ and is not stored in F-AP $m$, the contribution of other F-APs to F-AP $m$ is greater.
%Let $C_{n m}^{f}$ denote the contribution of F-AP $n$ to F-AP $m$ when content $f$ is delivered from F-AP $n$ to F-AP $m$.
%Then, it can be expressed as follows:
%\begin{equation}
%	C_{n m}^{f}=p_{m f}{L}\left(1-p_{n f}\right) P_{m}.
%\end{equation}
%Assume that F-AP is also selfish like a user to some extent, it is
%%is only willing to provide content caching services for local users, and
%not willing to provide extra cache space and power consumption for non-local users. Therefore, the contribution of F-AP $n$ to F-AP $m$ can also be regarded as the loss of social friendship between F-AP $m$ and F-AP $n$.
%Moreover, we define the transmission delay reduction by transmitting content $f$ from F-AP $n$ to local user $u$ of F-AP $m$ as the gain of social friendship for F-AP $m$ to F-AP $n$.
%Then, it can be expressed as follows:
%\begin{equation}
%	G_{m n}^{f}={L}\left({1}/{R}-{1}/{R_{m u}}-{1}/{R_{n u}}\right).
%\end{equation}
%\end{comment}
According to stochastic geometry theory, the position of user is modeled as an independent homogeneous Poisson distribution with the density $\lambda_u$.
Then, the probability of user $u$ contacting a potential partner can be expressed as follows:
\begin{equation}
p_{u,u^{\prime}} = 1 - \exp \left(- \lambda_u \pi {d_{u,u^{\prime}}}^{2}   \right),
\end{equation}
where $d_{u,u^{\prime}}$ denotes the distance between user $u$ and $u^{\prime}$.
Let $p_{m,n}$ denote the probability of the associated user in F-AP $m$ communicating with the associated user in F-AP $n$, which can be expressed as follows:
\begin{equation}\label{communi_prob}
p_{m,n} = \sum_{u \in \mathcal{U}_{m}}{\sum_{u^{\prime} \in \mathcal{U}_{n}}{p_{u,u^{\prime}}}}.
\end{equation}
Let $\boldsymbol{s}_{m}=\left[p_{m,1}, p_{m,2}, \ldots, p_{m,f}, \ldots, p_{m,F}  \right]^T$ denote the sequence of local content popularity in F-AP $m$.
Let $\rho_{m,n}$ denote the content popularity similarity of the associated users in F-AP $m$ and $n$, which can be expressed as follows:
\begin{equation}\label{prefer_sim}
\rho_{m,n}= \frac{\mathbb{E}\left[  \left( \boldsymbol{s}_{m}-\mu(s_{m})\right)^T  \left(  \boldsymbol{s}_{n}-\mu(s_{n})\right) \right]}   {\sigma\left( s_{m}\right) \sigma\left( s_{n}\right)},
\end{equation}
where $\mu\left(\cdot\right)$ denotes the mean of the local content popularity sequence, $\sigma(\cdot)$ the variance and $\mathbb{E}[\cdot]$ the expectation.
Let $g_{m,n}$ denote the social contribution between F-AP $m$ and $n$, which can be expressed as follows:
\begin{equation}
	g_{m,n}=p_{m,n} \rho_{m,n}.
\end{equation}
The social loss is defined as the caching and transmission overhead consumed by the cooperative transmission among F-APs. Then, it can be expressed as follows:
\begin{equation}\label{social_cost}
c_{m,n}= \sum_{f \in \mathcal{F}} \sum_{u \in \mathcal{U}_m} \left( p_{u,f}J_{\mathrm{c}} + \frac{P_{m}}{{R_{m,n}}} \right){L}.
\end{equation}

Meanwhile, considering the impact of distance on social relationship, we can define the social relationship of F-AP $m$ and F-AP $n$ as follows:
\begin{equation}
\psi_{m,n}=\left\{\begin{array}{ll}{ e^{-\frac{d_{m,n}}{d_{\mathrm{t h}}}}\left(g_{m,n}-\delta c_{m,n}\right),} & {d_{m,n} \leq  d_{\mathrm{t h}}}, \\
{0,} & d_{m,n} >  d_{\mathrm{th}}{,} \end{array}\right.
\end{equation}
where $\delta $ denotes the social parameter according to the relative importance between social gain and loss, $d_{m n}$ the distance between F-AP $m$ and F-AP $n$, and $d_{\mathrm{t h}}$ the distance threshold.
Since two F-APs can forward data for each other, we  believe that they are mutually trustworthy and the social preference is equivalent between F-AP $m$ and F-AP $n$. The mutual social preference between two F-APs can be expressed as follows:% \cite{ZhaoF}
\begin{equation}
	u_{m}(n)=u_{n}(m)=\left\{\begin{array}{ll}{ \psi_{m, n}+ \psi_{n, m},} & {m \neq n}, \\ {0,} & {m=n}{.}\end{array}\right.
\end{equation}
Obviously, the social preference is symmetric and additively separable \cite{Bogomonlaia}.
Correspondingly, F-AP $m$'s social preference for cluster $\mathcal{S}_{k}$ can be expressed as follows:
\begin{equation}
	U_{m}(\mathcal{S}_{k})=\sum_{n \in \mathcal{S}_{k}} u_{m}(n).
\end{equation}

\subsubsection{Proposed HCG-based Clustering}
Hedonic coalition game is known as a class of coalition formation game, which allows the formation of coalition by utilizing the individual preferences of the players \cite{Walid}.
If the coalition formation game is hedonic, the preference of any player depends only on the other members of the coalition that the player belongs to. Besides, the formation of the coalition is the consequence of the player's preference for possible coalition sets \cite{Bogomonlaia}.
The F-APs and clusters are equivalent to players and coalitions, respectively. Since the social preference of each F-AP for  cluster $\mathcal{S}_{k}$ depends entirely on local content popularity, distance and transmission power of other members in the same cluster, the formation of cluster can be modeled as a HCG.

Let $\Pi$ denote a coalition partition of the F-AP set $\mathcal{M}$, which is equivalent to the set of $K$ disjoint clusters.
Given a coalition partition $\Pi$ and a player $m$, let $\mathcal{S}_{\Pi}(m)$ denote the current cluster $\mathcal{S}_{k}$ to which F-AP $m$ belongs, and $\mathcal{S}_{\Pi}(m) \in \Pi$ such that $m \in \mathcal{S}_{\Pi}(m)$ \cite{Walid}. F-AP $m$'s social preference over possible clusters can be shown by an order $\succeq_m$, which is a reflexive, transitive and complete binary relation. Meanwhile, let $\succ_m$ denote the corresponding asymmetric relation.
If F-AP $m$ prefers to be a part of cluster $\mathcal{S}_{k}$ rather than cluster $\mathcal{S}_{k^{\prime}}$, the relationship can be established as follows:
\begin{equation}
	\mathcal{S}_{k}\succeq_m \mathcal{S}_{k^{\prime}} \Leftrightarrow \sum_{n \in \mathcal{S}_{k}} u_{m}(n)\geq\sum_{n \in  \mathcal{S}_{k^{\prime}}} u_{m}(n).
\end{equation}

In order to investigate the stability of hedonic game in the considered model, we introduce some concepts taken from \cite{Bogomonlaia}: 1) If there does not exist $m\in\mathcal{M}$ and a coalition $\mathcal{S}_{k} \in \Pi\cup\{\varnothing\}$ such that $\mathcal{S}_{k} \cup\{m\}\succ_m \mathcal{S}_{\Pi}(m)$ and $\mathcal{S}_{k}\cup\{m\}\succeq_n \mathcal{S}_{k}$ for all $n\in \mathcal{S}_{k}$, a coalition partition $\Pi$ is individually stable. 2) If the players' preferences are symmetric and additively separable, there exists a coalition partition which has individual stability. 3) Depending on the stability of the individual, if player $m$ can join the coalition without worsening the situation of any existing members, the coalition is considered to be open to player $m$, otherwise closed. Since the social preference is symmetric and additively separable, there exists an individually stable partition of cluster.

According to the above analysis, we propose a clustering scheme based on HCG.
Firstly, we randomly divide the F-AP set into several subsets. Then, we calculate the social preferences of F-AP $m$ for each cluster and find out $S$ clusters whose social preferences are greater than the current cluster $\mathcal{S}_{\Pi}(m)$. If these clusters are open to F-AP $m$, F-AP  $m$ is transferred into its preferable cluster. Repeat the above operations until the cluster partition is individually stable.
The detailed description of the proposed clustering scheme is shown in Algorithm \ref{alg-1}.

Let $P$ denote the maximum number of iterations when the partition is individually stable. Then, the computational complexity of the proposed HCG-based clustering scheme can be calculated to be $\mathcal{O}(PMK)$.
%Based on the social relationships among F-APs, the cached contents in F-APs will be more diverse. By using hedonic coalition game, F-APs with greater load difference will cooperate with each other.
%Besides, the formed clusters through this scheme can effectively improve the  cooperation gain among F-APs.

\begin{algorithm}[t]
	\renewcommand{\algorithmicrequire}{\textbf{Input:}}
	\renewcommand{\algorithmicensure}{\textbf{Output:}}
	\caption{Proposed HCG-based Clustering Scheme}
	\label{alg-1}
	\begin{algorithmic}[1]
		\Require the F-AP set ${\mathcal M}$;
		
		\State \textbf{Initialize:} Randomly divide ${\mathcal M}$ into several subsets;
		%\For {$m$ = 1 : $M$}
		\State Calculate F-AP $m$'s preference $u_{m}(n)$ with other F-APs;
		%\EndFor
		
		\Repeat
		\For {m = 1 : ${M}$}
		\State Calculate F-AP $m$'s preference $U_{m}({\mathcal{S}_{k}})$;
		\State Find clusters with preference greater than $\mathcal{S}_{\Pi}(m)$;
		\State Sort the obtained $S$ clusters in descending order;
		\For {$s$ = 1 : ${S}$}
		\If {$\mathcal{S}_{k}$ is open to F-AP $m$}
		\State $\mathcal{S}_{\Pi}(m) \rightarrow \mathcal{S}_{\Pi}(m) \backslash\{m\}$, $ \mathcal{S}_{k} \rightarrow \mathcal{S}_{k} \cup\{m\}$;
		%\State \textbf{break}
		\EndIf
		\EndFor
		\State \textbf{update} cluster partition $\Pi$;
		\EndFor
		\Until{Individual Stability;}
		\Ensure Individually stable coalition partition $\Pi$.
	\end{algorithmic}
\end{algorithm}

\subsection{Proposed Improved FA-based Cooperative Caching}
Since the optimization problem in (\ref{P-1}) is NP-hard, it can not be solved in polynomial time. On the other side,
the swarm intelligence algorithms do not rely on the analyticity of the optimization problem and can readily address the large-scale and complex problems which are difficult to be solved by brute force approaches \cite{Xia}. Hence, we propose a cooperative caching scheme through a swarm intelligence algorithm, i.e., the improved firefly algorithm, which can solve the optimization problem with lower computational complexity. Note that swarm intelligence algorithms can not always guarantee to find the global optimum. Our proposed improved firefly algorithm can increase the chance of convergence to the optimal solution.
%decrease the computational complexity and avoid pre-mature convergence.

%\subsection{Standard Firefly Algorithm}
The original FA was proposed in \cite{YangX} to mimic the process of firefly flash.
%The flashing light of firefly is utilized to attract mating partner or prey.
%By idealizing the characteristics of firefly flashing, for any two flashing fireflies, the darker one will fly towards the brighter one. The light intensity of a firefly is determined by the objective function value, and the brightness of firefly determines the attractiveness of firefly.
The position of each firefly represents a feasible solution of the optimization problem, and the brightest firefly is a better feasible solution to the optimization problem.
%The process of the original FA is briefly presented as follows. In the initialization phase, a population of fireflies will be randomly and uniformly initialized and the position of each firefly represents a feasible solution of the optimization problem. In the iteration phase, the location of each firefly is firstly evaluated by its light intensity, which is  proportional to the value of objective function. According to the brightness of different fireflies in the current iteration, the less brighter one will fly towards the brighter fireflies until the entire firefly population is traversed. Then the algorithm update the brightness of the fireflies and find the brightest firefly. The iterative process will terminate until the algorithm converges or reaches the maximal number of iteration. Finally, the brightest firefly is a better feasible solution to the optimization problem.
However, the original FA has some limitations. In the early stage of the iteration, if the brightness of a firefly greatly exceeds the average brightness of the current population, the firefly will quickly occupy an absolute proportion in the whole population. The diversity of the population rapidly decreases and the evolutionary ability will gradually disappear. Meanwhile, the search ability of FA depends mainly on the interaction between the firefly individuals, and the firefly lacks the corresponding mutation mechanism to get rid of the local extreme value.

%\subsection{Improved Firefly Algorithm}
To address the aforementioned issues, we propose an improved FA by utilizing the mutation strategy based on local content popularity.
The key in our proposed strategy is to meet the integer constraint by heaviside step function and avoid pre-mature convergence through the mutation strategy.

Let $\boldsymbol{X}^{i}=\left[x_{m,f}^{i}\right]_{M\times F}$ denote the $i$th firefly in each iteration, where $x_{m,f}^i$ denotes the $i$th caching status of content $f$ in F-AP $m$.
%where the $m$th row of the binary matrix denotes content caching status in F-AP $m$.
Let $\boldsymbol{X}^{\ast}$ denote the brightest individual firefly.
Here we simply use the objective function value of the firefly's location as the firefly's brightness. Let $I_i$ denote the brightness of the $i$th firefly. The attractiveness of firefly $i$ to adjacent firefly $j$ is associated with the brightness seen by adjacent firefly $j$, which varies with the distance between fireflies. Let $\beta_{i, j}$ denote the attractiveness of firefly $i$ to firefly $j$, which can be expressed as follows\cite{YangX}:
\begin{equation}\label{betaij}
	\beta_{i, j}=I_{i} e^{-\gamma r_{i, j}},
\end{equation}
where  $r_{i, j}=||\boldsymbol{X}^{i}-\boldsymbol{X}^{j}||_{1}$ is the manhattan distance between firefly $i$ and $j$, and $\gamma$ is the light absorption coefficient. In order to satisfy the integer constraint in (\ref{P-1}b), we propose to improve the method of firefly location update by exploiting heaviside step function.
Then, the movement of firefly $j$ attracted to the brighter firefly $i$ can be expressed as follows:
\begin{multline}\label{Move}
\boldsymbol{X}^{j}= \\
H\left(\boldsymbol{X}^{j}+\beta_{i, j}\left(\boldsymbol{X}^{i}-\boldsymbol{X}^{j}\right)+\lambda\left(\epsilon-\frac{1}{2}\right)-\frac{1}{2}\right),
\end{multline}
where $\lambda$ is a randomization parameter, $\epsilon$ is a uniformly distributed random number which is between 0 and 1, and $H(x)=\left\{\begin{array}{l}{0, x<0} \\ {1, x \geq 0}\end{array}\right.$ is the heaviside step function.% \cite{HuYa}.

%\vspace*{10pt}
\begin{algorithm}[t]
	\renewcommand{\algorithmicrequire}{\textbf{Input:}}
	\renewcommand{\algorithmicensure}{\textbf{Output:}}
	\caption{Proposed Improved FA-based Cooperative Caching Scheme}
	\label{alg-3}
	\begin{algorithmic}[1]
		
		\State \textbf{Initialize:} Generate initial population of fireflies;
		\While {not terminated}
		\State Evaluate individual light intensity $I_{i}$ ;
		\For {$j$ = 1 : $G$}
		
		\For {$i$ = 1 : ${G}$}
		
		\If {$I_{j}<I_{i}$}
		
		\State Calculate $\beta_{i, j}$ through (\ref{betaij});
		%\State Move firefly $j$ towards firefly $i$ through (\ref{Move});
		\State Calculate $\boldsymbol{X}^{j}$ through (\ref{Move});
		\EndIf
		\EndFor
		\If { $\sum_{f=1}^{F} x_{m, f} {L} > C$,}
		
		\State Rank stored contents by local popularity;
		\State Remove redundant and unpopular content;
		\Else \textbf{if}  { $\sum_{f=1}^{F} x_{m, f} {L} < C$,} \textbf{then}
		\State Calculate the remaining storage space;
		\State Cache popular content;
		
		\EndIf
		\EndFor	
		\EndWhile	
		\Ensure The brightest firefly $\boldsymbol{X}^{\ast}$.
		
	\end{algorithmic}
\end{algorithm}

Once firefly falls into the neighborhood of the extreme point, it is difficult to jump out. To solve this issue, we propose to adopt the mutation strategy based on local content popularity. When a firefly flies to a new location according to (\ref{Move}), the caching status is likely to slightly exceed or be less than the storage capacity of F-AP $m$. If the content caching status in F-AP $m$ slightly exceeds its storage capacity, the cached contents are sorted according to the local content popularity and the content with the lowest local content popularity will be deleted, i.e., $x_{m, f}^{i}=1$ turns into $x_{m, f}^{i}=0$. If the content caching status in F-AP $m$ is less than its storage capacity, all contents are sorted according to the local content popularity and the uncached content with the highest local content popularity will be stored, i.e., $x_{m, f}^{i}=0$ turns into $x_{m, f}^{i}=1$. Through the above mutation strategy, the capacity constraint in (\ref{P-1}a) can be satisfied. Meanwhile, the firefly can also easily jump out of the current local optimum and fly towards a better local optimum more quickly.
The detailed description of the improved FA-based cooperative caching scheme is shown in Algorithm \ref{alg-3}.

Let $Q$ denote the maximum number of iterations,
and $G$ the size of firefly population.
For the optimization problem in (\ref{P-1}), the computational complexity of the exhaustive search approach is $\mathcal{O}(2^{M F})$. In contrast, the computational complexity of the improved FA-based cooperative caching scheme is $\mathcal{O}(Q G^2)$. Generally, we have $F \gg G$, $F \gg Q$ and $M>2$. Correspondingly, $Q G^2 \ll 2^{M F}$ can always be guaranteed. Therefore, our proposed scheme has a significantly lower computational complexity than the traditional exhaustive search approach.

\section{Simulation Results}

In this section, the performance of the proposed cooperative caching scheme is evaluated via simulations. We consider a square area with side length of 1000 m and assume that the probability of user $u$ requesting content $f$ obeys Zipf distribution, i.e., $p_{u f}=\frac{R_{f}^{-\eta}}{\sum_{f' \in \mathcal{F}} R_{f'}^{-\eta}}$, where $\eta$ denotes the skewness factor and $R_{f}$ the rank of content $f$'s popularity in a descending order\cite{MML}.
The other system parameters are set as follows: $M=15$, $U=150$, $F=1000$, ${L}=500$ $\mathrm{MB}$, $B_{1}=B_{2}=10$ $\mathrm{MHz}$, $\sigma^{2}=-100$  $\mathrm{dBm}$,  $P_{m}=P_{\mathrm{s}}=46$ $\mathrm{dBm}$,  $\alpha=4$, $J_{\mathrm{c}}=6.25$ $\mathrm{pW/bit}$ \cite{LiuD},  $\mu=0.01$,  $\gamma=0.001$. We choose the random caching (RC) scheme and PIO-based cooperative caching scheme \cite{Xia} as the baselines for performance comparison.

\begin{figure}[t]
	\centering %\vspace*{135pt}
	\includegraphics[height=1.5in]{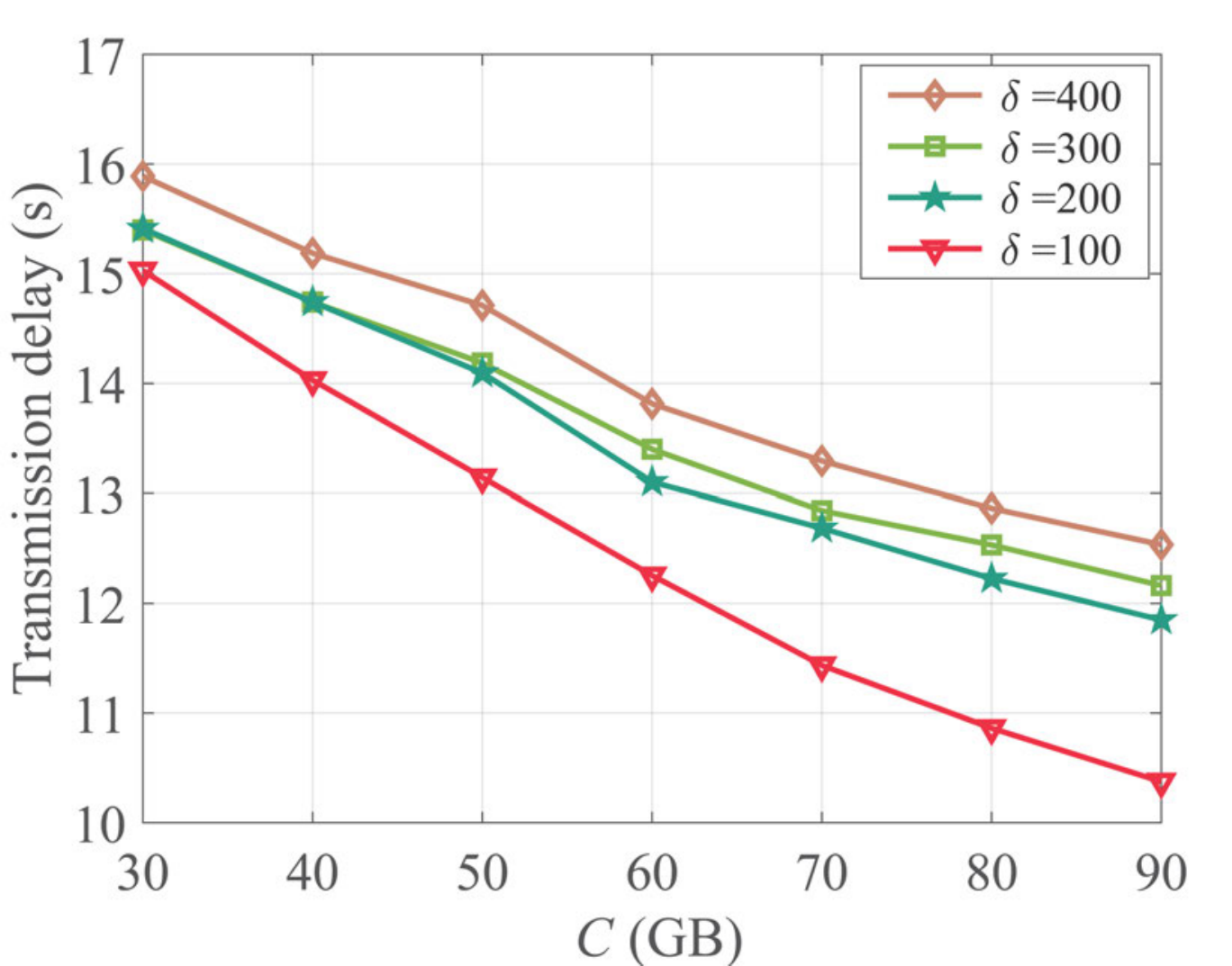}
	%\captionstyle{mystyle3}
	\caption{Transmission delay vs. storage capacity with different $\delta$.}
	\label{SD}
\end{figure}

%\vspace*{5pt}
\begin{figure*}[!t]
\begin{minipage}{0.28\textwidth}
\includegraphics[width=2.0in,height=1.5in]{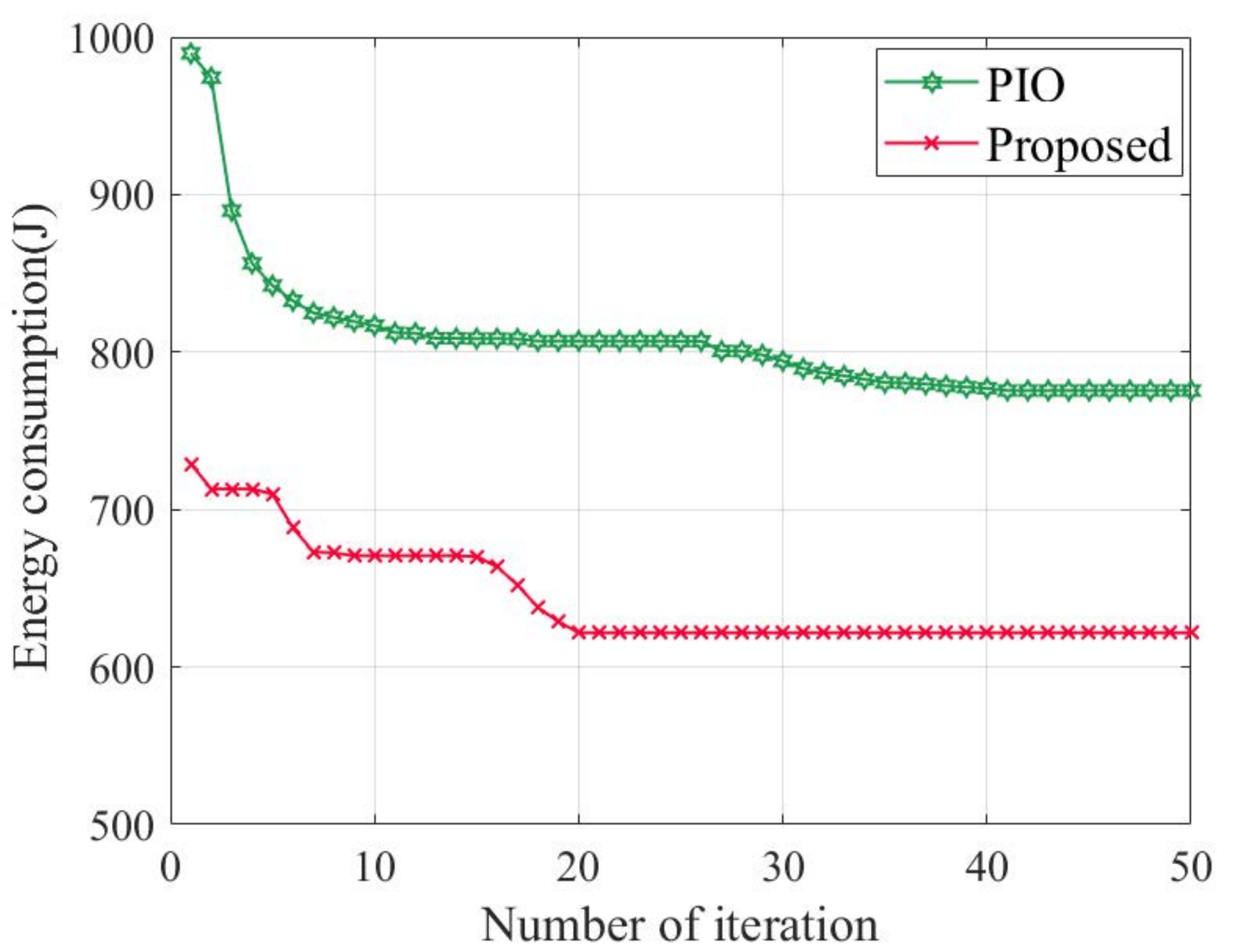}
\caption{Energy consumption vs. number of iterations.}
			\label{CVG}
\end{minipage}
\hfill
\begin{minipage}{0.28\textwidth}
\includegraphics[width=2.0in,height=1.5in]{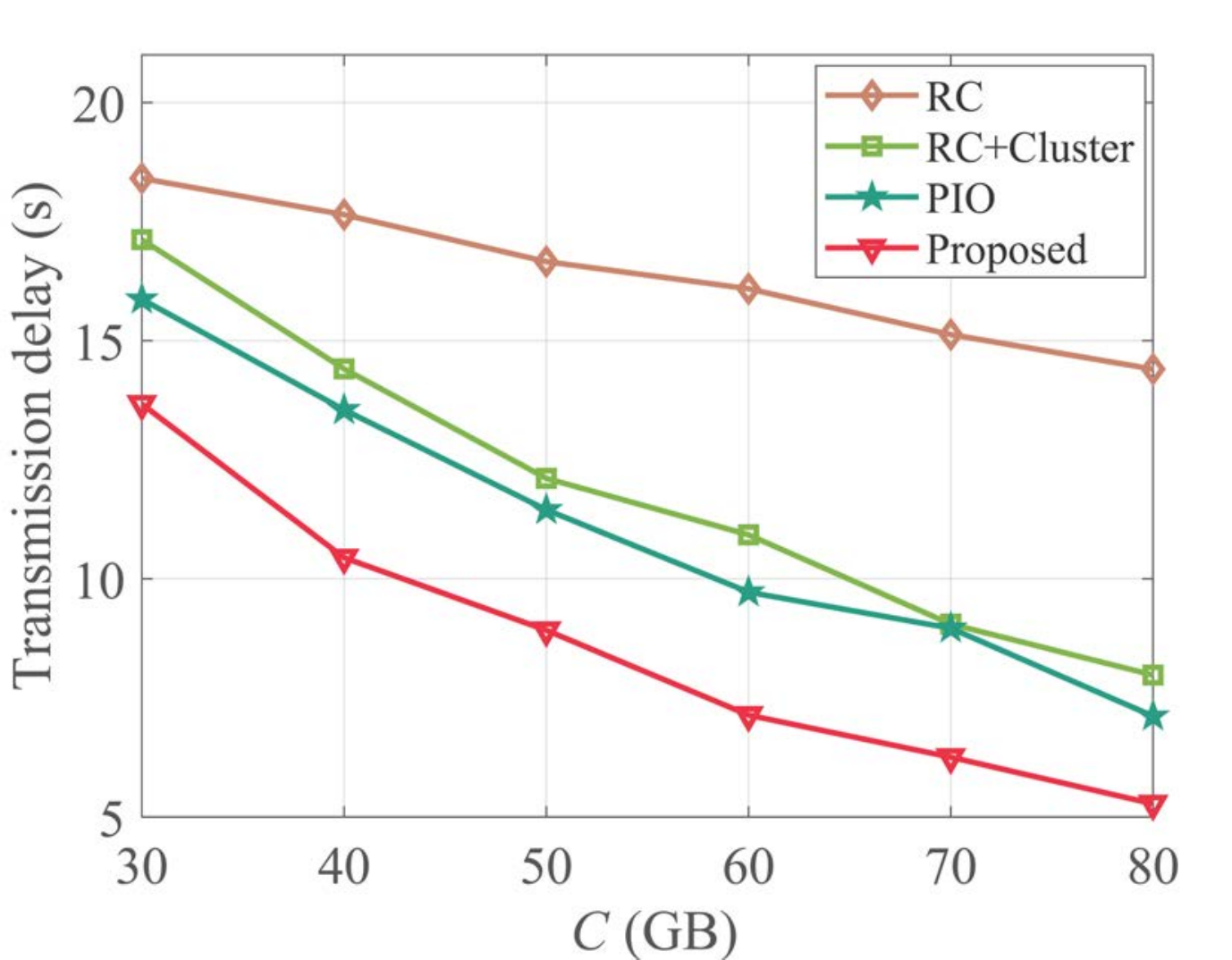}
\caption{Transmission delay vs. storage capacity with $\eta=0.5$.}
			\label{TD}
\end{minipage}
\hfill
\begin{minipage}{0.28\textwidth}
\includegraphics[width=2.0in,height=1.5in]{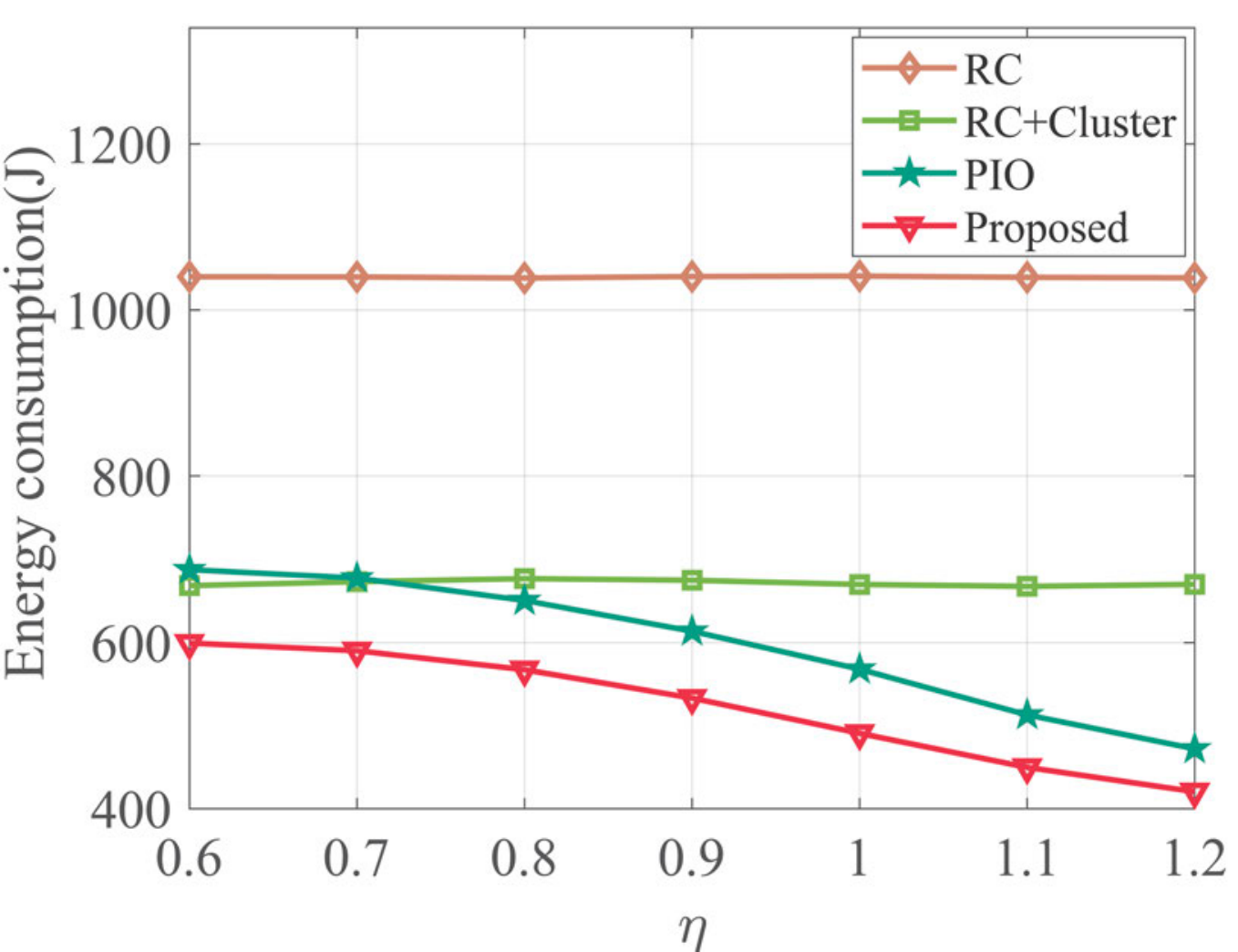}
\caption{Energy consumption vs. skewness factor with $C = 50$  $\mathrm{GB}$.}
			\label{EC}
\end{minipage}
\end{figure*}

In Fig. \ref{SD}, we show the transmission delay of our proposed scheme versus storage capacity $C$ with different social parameter $\delta$. It can be observed that the transmission delay increases with $\delta$. The reason is that when $\delta$ decreases, the proportion of social loss in social relationship is relatively decreased. Therefore, F-APs are more likely to cooperate with each other, and it is beneficial to consider social interaction among F-APs or clusters.

In Fig. \ref{CVG}, we show the convergence performance of our proposed scheme and the PIO-based scheme with $C = 50$  $\mathrm{GB}$ and $\eta=0.5$. It can be observed that compared with the PIO-based scheme, the improved FA can converge to a near optimum with a faster speed. The reason is that based on local content popularity, our proposed mutation strategy improves the searching ability of FA.
%It can also be observed that when only improved FA is considered, the energy consumption is smaller and better than PIO. The reason is that without considering the local content popularity, the lower hit rate of PIO results in extra network overhead.

In Fig. \ref{TD}, we show the transmission delay of our proposed scheme and the baselines with different storage capacity $C$. It can be observed that the transmission delay of all the considered schemes decreases with storage capacity. The reason is that as the storage capacity becomes larger, the hit rate in the storage increases.
It can also be observed that the proposed scheme is superior to the PIO-based scheme. The reason is that our proposed scheme can increase the cooperation gain among F-APs by exploiting the hidden social relationships.
%If we only consider the combination of clustering approach based on hedonic game and RC, our proposed clustering method can produce greater cooperation gains. Meanwhile, the proposed cooperative caching scheme is also superior to the PIO-based cooperative edge caching.

In Fig. \ref{EC},  we show the energy consumption of our proposed scheme and the baselines with different skewness factor $\eta$.
It can be observed that our proposed scheme outperforms the PIO-based scheme and the RC scheme. The reason is that our formulated cooperative caching model allows cooperation among clusters, which can improve the content diversity in F-APs and
reduce the overhead of content transmission among F-APs.
%It can also be observed that the energy consumption of our proposed clustering scheme outperforms RC with different $\eta$.
%The reason is that the content diversity in F-APs is significantly improved and the overhead of content transmission among F-APs is effectively reduced.
%F-APs in formed clusters still adopt RC scheme, which is not associated with content popularity.

%it can also be observed that the energy consumption of all considered schemes decreases with cache capacity. Note that our proposed clustering method can still generate greater cooperation gains to reduce energy consumption. Moreover, the energy consumption of the proposed cooperative caching scheme is also less than the PIO-based cooperative edge caching. The reason is that our proposed scheme considers the hidden social relationships among F-APs and allows cooperation among clusters.

%\begin{figure*}[!t]
%	\centering
%	\begin{tabular}{cc}
%		\begin{minipage}[t]{2.2in}
%			\includegraphics[width=2.1in]{figures/CVG}
%			\caption{Energy consumption vs. number of iteration.}
%			\label{CVG}
%		\end{minipage}
%%	\hspace{1.8ex}
%		\begin{minipage}[t]{2.15in}
%			\includegraphics[width=2.1in]{figures/TD}
%				\caption{Transmission delay vs. storage capacity with $\eta=0.5$.}
%			\label{TD}
%		\end{minipage}
%		%\hspace{0ex?}?
%		\begin{minipage}[t]{2.2in}
%			\includegraphics[width=2.1in]{figures/EC}
%			\caption{Energy consumption vs. skewness factor with $C = 50$  $\mathrm{GB}$.}
%			\label{EC}
%		\end{minipage}
%		
%		
%	\end{tabular}
%\end{figure*}

\section{Conclusions}

In this paper, we have proposed a social-aware cooperative caching scheme in F-RANs.
Using the proposed HCG-based clustering scheme, we have reduced the transmission delay and energy consumption by increasing the cooperation gain among F-APs.
In addition, using the proposed improved FA-based cooperative caching scheme, we have avoided pre-mature convergence with low computational complexity.
%Specially, by utilizing the local content popularity based mutation strategy, our improved FA has effectively avoided pre-mature convergence.
By utilizing the social relationships among F-APs and cooperation among clusters, our proposed scheme has provided considerable performance improvement over the baselines.

\section*{Acknowledgments}
This work was supported in part by the National Key Research and Development Program under Grant 2021YFB2900300, the National Natural Science Foundation of China under grant 61971129, and the Shenzhen Science and Technology Program under Grant KQTD20190929172545139.

\bibliographystyle{IEEEtran}

\bibliography{manuscript-coopcaching-conf}

\begin{thebibliography}{10}
\providecommand{\url}[1]{#1}
\csname url@samestyle\endcsname
\providecommand{\newblock}{\relax}
\providecommand{\bibinfo}[2]{#2}
\providecommand{\BIBentrySTDinterwordspacing}{\spaceskip=0pt\relax}
\providecommand{\BIBentryALTinterwordstretchfactor}{4}
\providecommand{\BIBentryALTinterwordspacing}{\spaceskip=\fontdimen2\font plus
\BIBentryALTinterwordstretchfactor\fontdimen3\font minus
  \fontdimen4\font\relax}
\providecommand{\BIBforeignlanguage}[2]{{%
\expandafter\ifx\csname l@#1\endcsname\relax
\typeout{** WARNING: IEEEtran.bst: No hyphenation pattern has been}%
\typeout{** loaded for the language `#1'. Using the pattern for}%
\typeout{** the default language instead.}%
\else
\language=\csname l@#1\endcsname
\fi
#2}}
\providecommand{\BIBdecl}{\relax}
\BIBdecl

\bibitem{Peng}
M.~Peng, S.~Yan, K.~Zhang, and et~al., ``Fog-computing-based radio access
  networks: Issues and challenges,'' \emph{IEEE Netw.}, vol.~30, no.~4, pp.
  46--53, Jul. 2016.

\bibitem{8632748}
Y.~Jiang, W.~Huang, M.~Bennis, and F.-C. Zheng, ``Decentralized asynchronous
  coded caching design and performance analysis in fog radio access networks,''
  \emph{IEEE Transactions on Mobile Computing}, vol.~19, no.~3, pp. 540--551,
  Feb. 2020.

\bibitem{jiang2020analysis}
Y.~Jiang, A.~Peng, C.~Wan, Y.~Cui, X.~You, F.-C. Zheng, and S.~Jin, ``Analysis
  and optimization of cache-enabled fog radio access networks: Successful
  transmission probability, fractional offloaded traffic and delay,''
  \emph{IEEE Transactions on Vehicular Technology}, vol.~69, no.~5, pp.
  5219--5231, Mar. 2020.

\bibitem{peng2018performance}
A.~Peng, Y.~Jiang, M.~Bennis, F.-C. Zheng, and X.~You, ``Performance analysis
  and caching design in fog radio access networks,'' in \emph{2018 IEEE
  Globecom Workshops (GC Wkshps)}.\hskip 1em plus 0.5em minus 0.4em\relax IEEE,
  Dec. 2018, pp. 1--6.

\bibitem{feng2019content}
H.~Feng, Y.~Jiang, D.~Niyato, F.-C. Zheng, and X.~You, ``Content popularity
  prediction via deep learning in cache-enabled fog radio access networks,'' in
  \emph{2019 IEEE Global Communications Conference (GLOBECOM)}.\hskip 1em plus
  0.5em minus 0.4em\relax IEEE, Dec. 2019, pp. 1--6.

\bibitem{lu2019distributed}
L.~Lu, Y.~Jiang, M.~Bennis, Z.~Ding, F.-C. Zheng, and X.~You, ``Distributed
  edge caching via reinforcement learning in fog radio access networks,'' in
  \emph{2019 IEEE 89th Vehicular Technology Conference (VTC2019-Spring)}.\hskip
  1em plus 0.5em minus 0.4em\relax IEEE, Apr. 2019, pp. 1--6.

\bibitem{hu2018distributed}
Y.~Hu, Y.~Jiang, M.~Bennis, and F.-C. Zheng, ``Distributed edge caching in
  ultra-dense fog radio access networks: A mean field approach,'' in \emph{2018
  IEEE 88th Vehicular Technology Conference (VTC-Fall)}.\hskip 1em plus 0.5em
  minus 0.4em\relax IEEE, Aug. 2018, pp. 1--6.

\bibitem{Xia}
C.~Xia, Y.~Jiang, M.~Peng, and et~al., ``Cooperative edge caching in fog radio
  access networks: A pigeon inspired optimization approach,'' in \emph{Proc.
  IEEE GLOBECOM}, Dec. 2019, pp. 1--6.

\bibitem{Fang}
S.~Fang and P.~Fan, ``A cooperative caching algorithm for cluster-based
  vehicular content networks with vehicular caches,'' in \emph{Proc. IEEE
  GLOBECOM Wkshps.}, Dec. 2017, pp. 1--6.

\bibitem{cuix}
Y.~Jiang, X.~Cui, M.~Bennis, and et~al., ``Cooperative caching in fog radio
  access networks: {A} graph-based approach,'' \emph{IET Commun.}, vol.~13,
  no.~20, pp. 3519--3528, Nov. 2019.

\bibitem{ZhangShan}
S.~Zhang, P.~He, K.~Suto, and et~al., ``Cooperative edge caching in
  user-centric clustered mobile networks,'' \emph{IEEE Trans. Mobile Comput.},
  vol.~17, no.~8, pp. 1791--1805, Aug. 2018.

\bibitem{ZhongC}
C.~Zhong, M.~C. Gursoy, and S.~Velipasalar, ``Deep multi-agent reinforcement
  learning based cooperative edge caching in wireless networks,'' in
  \emph{Proc. IEEE ICC}, May 2019, pp. 1--6.

\bibitem{LiZ}
Z.~Li, J.~Chen, and Z.~Zhang, ``Socially aware caching in {D2D} enabled fog
  radio access networks,'' \emph{IEEE Access}, vol.~7, pp. 84\,293--84\,303,
  Jun. 2019.

\bibitem{LiuD}
D.~Liu and C.~Yang, ``Energy efficiency of downlink networks with caching at
  base stations,'' \emph{IEEE J. Sel. Areas Commun.}, vol.~34, no.~4, pp.
  907--922, Apr. 2016.

\bibitem{ZHANGX}
X.~Zhang, Y.~Li, Y.~Zhang, and et~al., ``Information caching strategy for cyber
  social computing based wireless networks,'' \emph{IEEE Trans. Emerging Topics
  Comput.}, vol.~5, no.~3, pp. 391--402, Jul. 2017.

\bibitem{LIY}
Y.~Li, Z.~Zhang, H.~Wang, and et~al., ``{SERS}: Social-aware energy-efficient
  relay selection in {D2D} communications,'' \emph{IEEE Trans. Veh. Technol.},
  vol.~67, no.~6, pp. 5331--5345, Jun. 2018.

\bibitem{Bogomonlaia}
A.~Bogomonlaia and M.~Jackson, ``The stability of hedonic coalition
  structures,'' \emph{Games and Economic Behavior}, vol.~38, pp. 201--230, Jan.
  2002.

\bibitem{Walid}
W.~Saad, Z.~Han, M.~Debbah, and et~al., ``Coalitional game theory for
  communication networks,'' \emph{IEEE Signal Processing Mag.}, vol.~26, no.~5,
  pp. 77--97, Sep. 2009.

\bibitem{YangX}
X.~Yang, ``Firefly algorithms for multimodal optimization,'' \emph{Stochastic
  Algorithms: Foundations and Applications}, pp. 169--178, 2009.

\bibitem{MML}
Y.~Jiang, M.~Ma, M.~Bennis, and et~al., ``User preference learning based edge
  caching for fog radio access network,'' \emph{IEEE Trans. Commun.}, vol.~67,
  no.~2, pp. 1268--1283, Feb. 2019.

\end{thebibliography}

\end{document}